\documentstyle[12pt]{article}

   \def\000{{\bf 0}}
   \def\111{{\bf 1}}
   
   \def\mmm{\mid }
   \def\rrr{\rangle }
   \def\lll{\langle }

\begin{document}
 \title{Speedup in quantum computation is
associated with attenuation of processing probability}
\author{K. Svozil\\
 {\small Institut f\"ur Theoretische Physik}  \\
  {\small University of Technology Vienna }     \\
  {\small Wiedner Hauptstra\ss e 8-10/136}    \\
  {\small A-1040 Vienna, Austria   }            \\
  {\small e-mail: svozil@tph.tuwien.ac.at}}
\maketitle

\begin{flushright}
{\scriptsize kraft.tex}
\end{flushright}

\begin{abstract}
{\em
Quantum coherence
allows the computation of an arbitrary number of distinct computational
paths in parallel.
Based on quantum parallelism it has been conjectured that exponential
or even larger speedups of computations are possible. Here it is shown
that, although
in principle correct, any speedup is accompanied by an associated
attenuation of detection rates. Thus, on the average, no effective
speedup is obtained relative to classical (nondeterministic) devices.
}
\end{abstract}
 \clearpage

Recent findings in quantum complexity theory suggest an
exponential speedup of discrete logarithms and factoring \cite{shor}
and the travelling salesman problem \cite{cerni}  with respect to
classical complexity. (Classically, factoring an $n$-digit number takes
at most $O(n^{\log \log n})$ computation step; the travelling salesman
problem is $NP$-complete).
At the heart of these types of speedups is quantum parallelism.
Roughly stated, quantum parallelism assures that a single
quantum bit, henceforth called {\em qbit,}
can ``branch
off'' into an arbitrary number
of coherent entangled qbits.
A typical physical realization of a qbit is a single field mode of a
photon (electron, neutron), with the empty and the one-photon state
$\mmm
\000 \rrr$ and $\mmm \111 \rrr$ representing the classical symbols
$\000$ and
$\111$,
respectively. The branching process into coherent beam paths can be
realized
by an array of beam splitters such as semitransparent mirrors or a
double slit.  A typical cascade of branching process into $n^k$
coherent beam paths is
described by a successive array of $k$ identical beam splitters with $n$
slots and vanishing relative phases
\begin{eqnarray}
\mmm s_0 \rrr &\rightarrow &
{1\over \sqrt{n}} \left(
\mmm s_0s_{11}\rrr   +
\mmm s_0s_{12}\rrr   +
\cdots               +
\mmm s_0s_{1n}\rrr
\right)
\quad ,
\\
{1\over \sqrt{n}}
\mmm s_0s_{11}\rrr
 &\rightarrow &
{1\over n} \left(
\mmm s_0s_{11}s_{21}\rrr   +
\mmm s_0s_{11}s_{22}\rrr   +
\cdots                     +
\mmm s_0s_{11}s_{2n}\rrr
\right)
\\
{1\over \sqrt{n}}
\mmm s_0s_{12}\rrr
 &\rightarrow &
{1\over n}        \left(
\mmm s_0s_{12}s_{21}\rrr   +
\mmm s_0s_{12}s_{22}\rrr   +
\cdots                     +
\mmm s_0s_{12}s_{2n}\rrr
\right)
\\
&\vdots & \nonumber
\\
{1\over n^{-\, (k-1)/2}}
\mmm s_0s_{1n}\cdots s_{(k-1)n}\rrr
 &\rightarrow &
{1\over n^{-\, k/2}}        \left(
\mmm s_0s_{1n}\cdots s_{kn}\rrr   +
\cdots                                      +
\mmm s_0s_{1n}\cdots s_{kn}\rrr
\right)
\; .
\end{eqnarray}
Notice that every beam splitter contributes a normalization factor of
$1/\sqrt{n}$ to the amplitude of the process.
The probability amplitude for a single
quantum in state $\mmm s_0\rrr$ to evolve into one particular beam path
$s_0s_{1i_1}s_{2i_2}s_{3i_3}\cdots s_{ki_k}$ therefore is
\begin{equation}
\lll s_0s_{1i_1}s_{2i_2}s_{3i_3}\cdots s_{ki_k}\mmm U\mmm
s_0\rrr
= n^{-\,k/2}
\quad ,
\label{e:prob}
\end{equation}
where
$U$ stands for the unitary evolution operator corresponding to the
array of beam splitters.

More generally,
any one of the
entangled qbits originating from the branching
process can be processed in parallel.
The beam path
$s_0s_{1i_1}s_{2i_2}s_{3i_3}\cdots s_{ki_k}$
can be interpreted as a {\em program code}
\cite{hamming,chaitin,calude,svozil}.
How many programs can be coded into one beam path?
Notice that, in order to maintain coherence, no code of a valid program
can be the prefix of a code of another valid program. Therefeore, in
order to maintain the parallel quality of quantum computation,
only
{\em prefix} or {\em instantaneous} codes are allowed.
A straightforward proof using induction
\cite{hamming}
shows that
the instantaneous code
 of $q$ programs
 $\{p_1,p_2,\ldots ,p_q\}$ with length
 $l_1\le l_2\le \cdots \le l_q$ satisfies the {\em Kraft inequality}
 \begin{equation}
 \sum_{i=1}^q n^{-l_i}\le 1\quad ,
 \label{kraft}
 \end{equation}
 where $n$ is the number of symbols of the
code alphabet. In our case, $n$ is identified with the number of slits
in the beam splitters.
Stated pointedly, instantaneous decodability restricts the number of
legal programs due to the condition that to legal program can be the
prefix of another legal program.
The Kraft inequality then states that no more than maximally
$q=n^k$ programs can be coded by
a successive array of $k$ identical beam splitters with $n$
slots, corresponding to
 $l_1 = l_2 = \cdots = l_q$.
The more general case
$l_1\le l_2\le \cdots \le l_q$ can be easily
realized by allowing beams not to pass {\em all} $k$ $n$-slit arrays.

By recalling equation (\ref{e:prob}), it is easy to compute the
probability that a particular program $p_j$ of length $l_j\le k$
is executed. It is
 \begin{equation}
\vert \lll s_0s_{1i_1}s_{2i_2}s_{3i_3}\cdots s_{l_j\,i_{l_j}}\mmm U\mmm
s_0\rrr \vert^2
= n^{-\,{l_j}}
\quad .
 \end{equation}
Therefore, there is an inevitable exponential decrease
$ n^{-\,{l_j}} $
in the execution probability.

One possible way to circumvent attenuation would be to
amplify the output signals from the beam splitter array.
Classically, amplification and copying of bits is no big deal.
In quantum mechanics, however,
the no-cloning theorem \cite{no-cloning}
does not allow copying of quantum bits. Any attempt to copy qbits would
result in the addition of noise (e.g., from spontaneous emmission
processes) and, therefore, in errornous computations.

In summary, the price for
speedups of computations originating in
quantum parallelism is a corresponding attenuation of the computation
probability.
In order to compensate for an exponential decrease of execution
probability, one would have to
{\em exponentially increase} the number of (bosonic) quanta in the beam
paths. This, however, is equivalent to the trivial solution of an
arbitrarily complex
problem by the introduction of an arbitrary number of classical parallel
computers.


\begin{thebibliography}{99}

\bibitem{shor}
P. W. Shor, {\sl Algorithms for quantum computation: discrete logarithms
and factoring}, in {\sl Proc. 35th Annual Symposium of on Foundations of
Computer Science} (IEEE Press, November 1994), in press.


\bibitem{cerni}
V. \v{C}ern\'{y},
{\sl Phys. Rev.} {\bf A 48}, 116 (1993).

 \bibitem{hamming}
 R. W. Hamming, {\sl Coding and Information Theory, Second Edition}
 (Prentice-Hall, Englewood Cliffs, New Jersey, 1980).

\bibitem{chaitin}
G. J. Chaitin, {\sl Information, Randomness and Incompleteness, Second
edition}
(World Scientific, Singapore, 1987, 1990);
{\sl Algorithmic Information Theory}
(Cambridge University Press, Cambridge, 1987);
{\sl Information-Theoretic Incompleteness}
(World Scientific, Singapore, 1992).

\bibitem{calude}
C. Calude,
{\sl Information and Randomness --- An Algorithmic Perspective}
(Springer, Berlin, 1994).

\bibitem{svozil}
K. Svozil,  {\sl Randomness and Undecidability in Physics}
(World Scientific, Singapore, 1993).

\bibitem{no-cloning}
W. K. Wooters and W. H. Zurek,
{\sl Nature} {\bf 299}, 802 (1982);
L. Mandel,
{\sl Nature} {\bf 304}, 188 (1983);
P. W. Milonni and M. L. Hardies,
{\sl Phys. Lett.} {\bf 92A}, 321 (1982);
 R. J. Glauber, {\sl Amplifiers, Attenuators and the Quantum Theory of
 Measurement}, in {\sl Frontiers in Quantum Optics}, ed. by E. R. Pikes
 and S. Sarkar (Adam Hilger, Bristol 1986);
C. M. Caves, {\sl Phys. Rev.} {\bf D 26}, 1817 (1982).

\end{thebibliography}
\end{document}